\documentclass[showpacs,preprintnumbers,amsmath,amssymb,nofootinbib]{revtex4-1}

\usepackage{graphicx}
\usepackage{bm}

\usepackage{amssymb}
\usepackage{amsfonts}
\usepackage{amsmath}

\usepackage{latexsym}

\usepackage{dsfont}
\usepackage{multirow}

\usepackage{color}
\usepackage[mathscr]{eucal}

\definecolor{nv}{rgb}{0.1,0.1,0.6}
\definecolor{pr}{rgb}{0.2,0.1,0.5}
\definecolor{mg}{rgb}{0.4,0.0,0.4}

\newcommand{\nn}{\nonumber}

\newcommand{\beq}{\begin{equation}}
\newcommand{\eeq}{\end{equation}}
\newcommand{\beqy}{\begin{eqnarray}}
\newcommand{\eeqy}{\end{eqnarray}}
\newcommand{\beqyn}{\begin{eqnarray*}}
\newcommand{\eeqyn}{\end{eqnarray*}}
\newcommand{\nl}{\newline}

\newcommand{\bs}{\begin{slide}}
\newcommand{\es}{\end{slide}}
\newcommand{\bc}{\begin{center}}
\newcommand{\ec}{\end{center}}
\newcommand{\bmin}{\begin{minipage}}
\newcommand{\emin}{\end{minipage}}

\newcommand{\bi}{\begin{itemize}}
\newcommand{\ei}{\end{itemize}}

\newcommand{\la}{\langle}
\newcommand{\ra}{\rangle}

\begin{document}


\title{The end of WHAT nucleon-spin crisis?}

\author{Elliot Leader}
\email{e.leader@imperial.ac.uk}
\affiliation{Imperial College London\\ Prince Consort Road, London SW7 2AZ }

\date{\today}

\begin{abstract}

Povh and Walcher have written a paper entitled ``The end of the nucleon-spin crisis". But there is no such crisis. What appeared to be a spin crisis in the parton model, 28 years ago, was a consequence of a misinterpretation of the results of
the famous European Muon Collaboration experiment on polarized deep inelastic scattering. It would thus seem that Povh and Walcher have invoked a somewhat dubious argument based on hadronic fluctuations of the photon in order to resolve a non-existent problem.

\end{abstract}


\maketitle


\section{Introduction\label{secI}}
In a recent paper Povh and Walcher \cite{Povh:2016kvg} offer a solution to what  they call ``the nucleon-spin crisis",  referring to the1988 European Muon Collaboration (EMC) experiment  \cite{Ashman:1987hv, Ashman:1989ig} on fully inclusive polarized Deep Inelastic Scattering (DIS), which \emph{at the time} was interpreted as implying a problem with understanding the spin of the nucleon, and which gave rise to a paper \cite{Leader:1988vd} with the enigmatic title \emph{A crisis in the parton model: where, oh where is the proton's spin?}. However, it has long been understood that there is no such crisis and  that the belief in a  spin-crisis emerged from an over  naive interpretation of the EMC experiment. For a modern summary of the situation and access to the literature, see the review \cite{Kuhn:2008sy}.\nl

\section{The naive interpretation of the EMC results}
Before the invention of QCD, in what was called the quark model of the nucleon, strictly,  the \emph{constituent} quark model, a nucleon (N) is visualized as a bound state  of 3 massive quarks (Q) ( $ M_Q \approx  M_N/3 $)  lying in some kind of potential. The nucleon corresponds to the ground state, and for any reasonably behaved potential this will be an s-state. Models   of various degrees of sophistication give results in good agreement with the static properties of the nucleon. In the simplest non-relativistic case, for an s-state the constituent quarks have no orbital angular momentum (OAM) and one has, for the nucleon at rest, say polarized in the positive  Z-direction
\beq \label{naive-spin} 1/2 =   S_z^{N}= \sum_Q S_z^Q.\eeq
With relativistic corrections, the bottom  components of the quark Dirac spinors contain OAM and Eq.~(\ref{naive-spin}) is modifed to
\beq \label{rel-naive} \sum_Q S_z^Q \approx 0.3 \eeq
Let us now see how the EMC results were  misconstrued as being in significant disagreement with both (\ref{naive-spin}) and ( \ref{rel-naive}).

\section{Rigorous interpretation of polarized DIS experimental results}
The QCD Lagrangian is expressed in terms of quark and gluon fields. The quanta of these quark fields (strictly these are \emph{partonic}  quarks, also referred to as current quarks) do \textbf{not} correspond to the constituent quarks discussed above. Moreover the connection  between these two types of quarks is unknown. Such a relationship would depend on highly non-perturbative aspects of QCD.\nl
In considering deep inelastic scattering on a proton the quantity of interest is the first moment of the spin-dependent structure function $g_1(x,Q^2)$ :
\begin{equation}\label{eq:Gamma1}
  \Gamma _1^p \equiv \int_0^1 g_1^p(x) dx \eeq
  which can be expressed in terms of the proton matrix elements of the Gell-Mann flavour octet  of axial-vector currents,
  \beq \label{eq:AVcurrent}
J^i_{5\mu} = {\bar{\bm{\psi}}} \gamma_{\mu}\gamma_5
\Biggl(\frac{{\bm{\lambda}}_i }{2}\Biggl) {\bm{\psi}}\qquad
(i=1,2,...,8),
\eeq
where the ${\bm{\lambda}}_j$ are the usual Gell-Mann matrices and
${\bm{\psi}}$ is a column vector in flavor space
\begin{equation}\label{eq:psi}
{\bm{\psi}} =
\left(\begin{array}{c}\psi_u\\\psi_d\\\psi_s\end{array} \right)
\end{equation}
and a  flavor singlet current
\begin{equation}\label{eq:AVsinglet}
J^0_{5\mu} = {\bar{\bm{\psi}}} \gamma_\mu \gamma_5 {\bm{\psi}}\,.
\end{equation}
The matrix elements have the form
\beqy \label{eq:defain}
\langle P,S|J^3_{5\mu}|P,S\rangle & =& Ma_3S_\mu  \nn \\
\langle P,S|J^8_{5\mu}|P,S\rangle & = & M\,\,a_8S_\mu  \nn \\
\langle P,S|J^0_{5\mu}|P,S\rangle &=& 2M a_0(Q^2) S_\mu . \eeqy
where the $Q^2$ dependence of $a_0$ arises because the singlet current has to be renormalized, and it is customary and convenient to choose $ Q^2 $ as the renormalization scale, whereas, to the extent that $SU(3)_{\textrm{flavour}}$ is a good symmetry so that the octet of curents is conserved, $a_3$ and $a_8$ are independent of $Q^2$. \nl
In NLO the expression for $\Gamma _1^p$ in terms of these is scheme dependent. In the $\overline{MS}$ scheme,
at leading twist, valid for $Q^2\gg M^2$,
\beq \label{eq:GammaTotal}
\Gamma_1^{p}(Q^2) = \frac{1}{12}\big[ \big( a_3 + \frac{1}{3}\,a_8) \, \Delta C_{NS}^{\overline{MS}} + \frac{4}{3}\,a_0(Q^2)\,\Delta C_{S}^{\overline{MS}} \big] , \eeq
where the $\Delta C$ are the known singlet and non-singlet Wilson coefficients. \nl
Now the values of $a_{3,8}$ can be obtained from neutron and hyperon  $\beta$-decay. This follows from the Wigner-Eckart theorem on the assumption that $SU(3)_{\textrm{flavour}}$ is a good symmetry and that the octet of baryons transform as a flavour octet. \nl
\emph{Hence a mesurement of $\Gamma_1^{p}$ at some value of  $\, Q^2$ is effectively a measure of} $a_0(Q^2)$.

\section{Misinterpretation of the measured value of $a_0$ }
In the \emph{naive} parton model, for a fast moving proton with helicity $+1/2$,
\beq \label{simple} a_0= \Delta \Sigma \equiv (\Delta u + \Delta\overline{u}) + (\Delta d + \Delta\overline{d}) + (\Delta s + \Delta\overline{s})
\end{equation}
where the $\Delta q $, $\Delta \bar{q}$ are the first moments of the polarized quark-parton helicity densities. Hence, bearing in mind that $\Delta q = q_{+} -q_{-} $, where $\pm $ corresponds to the number densities with spin along or opposite to the proton's momentum, so that the average spin carried by a given quark is given by $\langle  S_z^{q} \rangle = 1/2 \,\Delta q $, one obtains  in the naive parton model
\beq a_0 = \Delta \Sigma =2 \left[\sum_{q} \langle  S_z^{q} \rangle + \sum_{\bar{q}} \langle  S_z^{\bar{q}} \rangle\right] \eeq
and if there is no other source of angular momentum one expects
\beq \label{wrong} \left[\sum_{q} \langle  S_z^{q} \rangle + \sum_{\bar{q}} \langle  S_z^{\bar{q}} \rangle\right]= S_z^{\textrm{proton}}= 1/2 \eeq
implying, naively,
\beq \label{nivespin} a_0 = 1. \eeq
The EMC experiment gave $a_0\approx 0 $ and later experiments confirmed that $a_0 \ll 1 $, giving rise to the spin crisis in the (\emph{naive}) parton model.\nl
However,  Eq.~(\ref{nivespin}) \emph{ cannot possibly be true} because the right hand side is a fixed number, whereas the left hand side is, \emph{beyond the naive level}, equal to $a_0(Q^2)$, i.e. a function of $Q^2$! Thus failure of Eq.~(\ref{nivespin}) to hold cannot possibly be used to infer that there is  crisis\footnote{Note that beyond the naive level, in the $\overline{MS}$ scheme, Eq.~(\ref{simple}) still holds, but both sides of the equation then depend on $Q^2$.}.\nl
It is obvious that a correct relation between the spin of a  nucleon and the angular momentum of its constituents should include their orbital angular momentum and should also include a contribution from the gluons. Unfortunately this is much more complicated than it sounds, because there is some controversy as to which operators  should be  used to represent the angular momentum, especially in the case of the massless gluon \cite{Leader:2011za,Leader:2013jra}. The relation, based on the canonical  (can) version of the angular momentum,
 \beq \label{srule}
\tfrac{1}{2}= \la \la \hat{S}^q_z  \ra \ra + \la \la  \hat{L}^q_z  \ra \ra + \la \la  \hat{S}^G_z \ra \ra  + \la \la  \hat{L}^G_z  \ra \ra
\eeq
 looks totally intuitive; can't be incorrect.
Usually  Eq.~(\ref{srule}) is written in  the Jaffe-Manohar (JM) form \cite{Jaffe:1989jz}:
\beq
\tfrac{1}{2}= \tfrac{1}{2} a_0  +  \Delta G  + \la \la \hat{ L}^q_z  \ra \ra  + \la \la  \hat{L}^G_z  \ra \ra
\eeq
but more correctly it should read :
\beq
\tfrac{1}{2}= \tfrac{1}{2} a_0  +  \Delta G  + \la \la  \hat{L}^q_{can,z}  \ra \ra  + \la \la  \hat{L}^G_{can,z}  \ra \ra .
\eeq
But this is still not completely accurate. Danger! $\Delta G $ is a gauge invariant quantity but $  \la \la  S^G_{\textrm{can},z} \ra \ra $ is  not. \nl
 However one can show that \cite{Anselmino:1994gn}
 \beq \Delta G  =  \la \la \hat{ S}^G_{\textrm{can},z} \ra \ra \large|_{\textrm{Gauge} A^0=0}, \eeq
or,  as the nucleon momentum $P\rightarrow \infty $
\beq \Delta G  =  \la \la \hat{ S}^G_{\textrm{can},z} \ra \ra \large|_{\textrm{Gauge} A^+=0} . \eeq
  Moreover the operators $\hat{L}^{q,G}$  are also not gauge invariant. Thus all the gauge non-invariant operators appearing  in the JM sum rule should be evaluated in the gauge $ A^0=0 $ or $ A^+=0 $. Finally then, for a fast moving proton with helicity $+1/2$ the angular momentum sum rule becomes, in contrast to the naive result Eq.~(\ref{nivespin}),
   \beq \label{correct} \tfrac{1}{2}= \tfrac{1}{2} a_0  +  \Delta G  + \la \la  \hat{L}^q_{can,z}  \ra \ra\large|_{A^+=0}  + \la \la  \hat{L}^G_{can,z}  \ra \ra\large|_{A^+=0}.\eeq
 It should  not be forgotten that each individual term in Eq.~(\ref{correct}) is actually a function of $Q^2$, but that the sum is not.\nl
 $ \Delta G $ can be measured and seems to be relatively small, but not negligible,  typically  $\Delta G \approx 0.29 \pm 0.32 $ for  $Q^2 \approx 10GeV^2 $ \cite{Kuhn:2008sy} (see also \cite{PhysRevLett.113.012001}). The real challenge is to find a way to measure the orbital terms in Eq.~(\ref{correct}) and thus to check whether the sum rule holds.

\section{Conclusion}
What appeared to be a spin crisis in the parton model, 28 years ago, was a consequence of a misinterpretation of the results of
the famous European Muon Collaboration experiment on polarized deep inelastic scattering. This was caused by a failure to distinguish adequately between constituent and partonic quarks. In constituent quark models of the nucleon, in say a proton with ``spin up", the sum of the Z-components of the spins of the \emph{constituent} quarks, depending on the sophistication of the model, is expected to be close to $+1/2$. The European Muon Collaboration, and later experiments, confirmed that the sum of the Z-components of the \emph{partonic} quarks was very different, much smaller than $1/2$. But we do not understand the connection between partonic and constituent quarks---a non-perturbative dynamical question---so that there is no basis for regarding this difference as a crisis. Moreover the smallness of the spin contribution of the partonic quarks is perfectly reasonable, given that they certainly  possess orbital angular momentum as well, and that the gluons, too, carry some spin. Thus there simply isn't a nucleon-spin crisis, and the paper of Povh and Walcher appears to be an attempt to resolve a non-existent problem.

\section{Acknowledgement} I thank Enrico Predazzi and Dimiter Stamenov for helpful suggestions.

\bibliography{Elliot_General}

\begin{thebibliography}{10}%
\makeatletter
\providecommand \@ifxundefined [1]{%
 \@ifx{#1\undefined}
}%
\providecommand \@ifnum [1]{%
 \ifnum #1\expandafter \@firstoftwo
 \else \expandafter \@secondoftwo
 \fi
}%
\providecommand \@ifx [1]{%
 \ifx #1\expandafter \@firstoftwo
 \else \expandafter \@secondoftwo
 \fi
}%
\providecommand \natexlab [1]{#1}%
\providecommand \enquote  [1]{``#1''}%
\providecommand \bibnamefont  [1]{#1}%
\providecommand \bibfnamefont [1]{#1}%
\providecommand \citenamefont [1]{#1}%
\providecommand \href@noop [0]{\@secondoftwo}%
\providecommand \href [0]{\begingroup \@sanitize@url \@href}%
\providecommand \@href[1]{\@@startlink{#1}\@@href}%
\providecommand \@@href[1]{\endgroup#1\@@endlink}%
\providecommand \@sanitize@url [0]{\catcode `\\12\catcode `\$12\catcode
  `\&12\catcode `\#12\catcode `\^12\catcode `\_12\catcode `\%12\relax}%
\providecommand \@@startlink[1]{}%
\providecommand \@@endlink[0]{}%
\providecommand \url  [0]{\begingroup\@sanitize@url \@url }%
\providecommand \@url [1]{\endgroup\@href {#1}{\urlprefix }}%
\providecommand \urlprefix  [0]{URL }%
\providecommand \Eprint [0]{\href }%
\providecommand \doibase [0]{http://dx.doi.org/}%
\providecommand \selectlanguage [0]{\@gobble}%
\providecommand \bibinfo  [0]{\@secondoftwo}%
\providecommand \bibfield  [0]{\@secondoftwo}%
\providecommand \translation [1]{[#1]}%
\providecommand \BibitemOpen [0]{}%
\providecommand \bibitemStop [0]{}%
\providecommand \bibitemNoStop [0]{.\EOS\space}%
\providecommand \EOS [0]{\spacefactor3000\relax}%
\providecommand \BibitemShut  [1]{\csname bibitem#1\endcsname}%
\let\auto@bib@innerbib\@empty
\bibitem [{\citenamefont {Povh}\ and\ \citenamefont
  {Walcher}(2016)}]{Povh:2016kvg}%
  \BibitemOpen
  \bibfield  {author} {\bibinfo {author} {\bibfnamefont {B.}~\bibnamefont
  {Povh}}\ and\ \bibinfo {author} {\bibfnamefont {T.}~\bibnamefont {Walcher}},\
  }\href@noop {} {\  (\bibinfo {year} {2016})},\ \Eprint
  {http://arxiv.org/abs/1603.05884} {arXiv:1603.05884 [hep-ph]} \BibitemShut
  {NoStop}%
\bibitem [{\citenamefont {Ashman}\ \emph {et~al.}(1988)\citenamefont {Ashman}
  \emph {et~al.}}]{Ashman:1987hv}%
  \BibitemOpen
  \bibfield  {author} {\bibinfo {author} {\bibfnamefont {J.}~\bibnamefont
  {Ashman}} \emph {et~al.} (\bibinfo {collaboration} {European Muon}),\
  }\bibfield  {booktitle} {\emph {\bibinfo {booktitle} {{Internal spin
  structure of the nucleon. Proceedings, Symposium, SMC Meeting, New Haven,
  USA, January 5-6, 1994}}},\ }\href {\doibase 10.1016/0370-2693(88)91523-7}
  {\bibfield  {journal} {\bibinfo  {journal} {Phys. Lett.}\ }\textbf {\bibinfo
  {volume} {B206}},\ \bibinfo {pages} {364} (\bibinfo {year}
  {1988})}\BibitemShut {NoStop}%
\bibitem [{\citenamefont {Ashman}\ \emph {et~al.}(1989)\citenamefont {Ashman}
  \emph {et~al.}}]{Ashman:1989ig}%
  \BibitemOpen
  \bibfield  {author} {\bibinfo {author} {\bibfnamefont {J.}~\bibnamefont
  {Ashman}} \emph {et~al.} (\bibinfo {collaboration} {European Muon}),\
  }\bibfield  {booktitle} {\emph {\bibinfo {booktitle} {{Internal spin
  structure of the nucleon. Proceedings, Symposium, SMC Meeting, New Haven,
  USA, January 5-6, 1994}}},\ }\href {\doibase 10.1016/0550-3213(89)90089-8}
  {\bibfield  {journal} {\bibinfo  {journal} {Nucl. Phys.}\ }\textbf {\bibinfo
  {volume} {B328}},\ \bibinfo {pages} {1} (\bibinfo {year} {1989})}\BibitemShut
  {NoStop}%
\bibitem [{\citenamefont {Leader}\ and\ \citenamefont
  {Anselmino}(1988)}]{Leader:1988vd}%
  \BibitemOpen
  \bibfield  {author} {\bibinfo {author} {\bibfnamefont {E.}~\bibnamefont
  {Leader}}\ and\ \bibinfo {author} {\bibfnamefont {M.}~\bibnamefont
  {Anselmino}},\ }\bibfield  {booktitle} {\emph {\bibinfo {booktitle}
  {{Proceedings, 8th International Symposium on High-energy Spin Physics}}},\
  }\href {\doibase 10.1063/1.38312, 10.1007/BF01566922} {\bibfield  {journal}
  {\bibinfo  {journal} {Z. Phys.}\ }\textbf {\bibinfo {volume} {C41}},\
  \bibinfo {pages} {239} (\bibinfo {year} {1988})}\BibitemShut {NoStop}%
\bibitem [{\citenamefont {Kuhn}\ \emph {et~al.}(2009)\citenamefont {Kuhn},
  \citenamefont {Chen},\ and\ \citenamefont {Leader}}]{Kuhn:2008sy}%
  \BibitemOpen
  \bibfield  {author} {\bibinfo {author} {\bibfnamefont {S.~E.}\ \bibnamefont
  {Kuhn}}, \bibinfo {author} {\bibfnamefont {J.~P.}\ \bibnamefont {Chen}}, \
  and\ \bibinfo {author} {\bibfnamefont {E.}~\bibnamefont {Leader}},\ }\href
  {\doibase 10.1016/j.ppnp.2009.02.001} {\bibfield  {journal} {\bibinfo
  {journal} {Prog. Part. Nucl. Phys.}\ }\textbf {\bibinfo {volume} {63}},\
  \bibinfo {pages} {1} (\bibinfo {year} {2009})},\ \Eprint
  {http://arxiv.org/abs/0812.3535} {arXiv:0812.3535 [hep-ph]} \BibitemShut
  {NoStop}%
\bibitem [{\citenamefont {Leader}(2011)}]{Leader:2011za}%
  \BibitemOpen
  \bibfield  {author} {\bibinfo {author} {\bibfnamefont {E.}~\bibnamefont
  {Leader}},\ }\href {\doibase 10.1103/PhysRevD.83.096012} {\bibfield
  {journal} {\bibinfo  {journal} {Phys.Rev.}\ }\textbf {\bibinfo {volume}
  {D83}},\ \bibinfo {pages} {096012} (\bibinfo {year} {2011})},\ \Eprint
  {http://arxiv.org/abs/1101.5956} {arXiv:1101.5956 [hep-ph]} \BibitemShut
  {NoStop}%
\bibitem [{\citenamefont {Leader}\ and\ \citenamefont
  {Lorce}(2013)}]{Leader:2013jra}%
  \BibitemOpen
  \bibfield  {author} {\bibinfo {author} {\bibfnamefont {E.}~\bibnamefont
  {Leader}}\ and\ \bibinfo {author} {\bibfnamefont {C.}~\bibnamefont {Lorce}},\
  }\href {\doibase 10.1016/j.physrep.2014.02.010} {\bibfield  {journal}
  {\bibinfo  {journal} {Phys.Rep.}\ }\textbf {\bibinfo {volume} {541}},\
  \bibinfo {pages} {163} (\bibinfo {year} {2013})},\ \Eprint
  {http://arxiv.org/abs/1309.4235} {arXiv:1309.4235 [hep-ph]} \BibitemShut
  {NoStop}%
\bibitem [{\citenamefont {Jaffe}\ and\ \citenamefont
  {Manohar}(1990)}]{Jaffe:1989jz}%
  \BibitemOpen
  \bibfield  {author} {\bibinfo {author} {\bibfnamefont {R.~L.}\ \bibnamefont
  {Jaffe}}\ and\ \bibinfo {author} {\bibfnamefont {A.}~\bibnamefont
  {Manohar}},\ }\href {\doibase 10.1016/0550-3213(90)90506-9} {\bibfield
  {journal} {\bibinfo  {journal} {Nucl. Phys.}\ }\textbf {\bibinfo {volume}
  {B337}},\ \bibinfo {pages} {509} (\bibinfo {year} {1990})}\BibitemShut
  {NoStop}%
\bibitem [{\citenamefont {Anselmino}\ \emph {et~al.}(1995)\citenamefont
  {Anselmino}, \citenamefont {Efremov},\ and\ \citenamefont
  {Leader}}]{Anselmino:1994gn}%
  \BibitemOpen
  \bibfield  {author} {\bibinfo {author} {\bibfnamefont {M.}~\bibnamefont
  {Anselmino}}, \bibinfo {author} {\bibfnamefont {A.}~\bibnamefont {Efremov}},
  \ and\ \bibinfo {author} {\bibfnamefont {E.}~\bibnamefont {Leader}},\
  }\href@noop {} {\bibfield  {journal} {\bibinfo  {journal} {Phys. Rep.}\
  }\textbf {\bibinfo {volume} {261}},\ \bibinfo {pages} {1} (\bibinfo {year}
  {1995})},\ \Eprint {http://arxiv.org/abs/hep-ph/9501369} {hep-ph/9501369}
  \BibitemShut {NoStop}%
\bibitem [{\citenamefont {de~Florian}\ \emph {et~al.}(2014)\citenamefont
  {de~Florian}, \citenamefont {Sassot}, \citenamefont {Stratmann},\ and\
  \citenamefont {Vogelsang}}]{PhysRevLett.113.012001}%
  \BibitemOpen
  \bibfield  {author} {\bibinfo {author} {\bibfnamefont {D.}~\bibnamefont
  {de~Florian}}, \bibinfo {author} {\bibfnamefont {R.}~\bibnamefont {Sassot}},
  \bibinfo {author} {\bibfnamefont {M.}~\bibnamefont {Stratmann}}, \ and\
  \bibinfo {author} {\bibfnamefont {W.}~\bibnamefont {Vogelsang}},\ }\href
  {\doibase 10.1103/PhysRevLett.113.012001} {\bibfield  {journal} {\bibinfo
  {journal} {Phys. Rev. Lett.}\ }\textbf {\bibinfo {volume} {113}},\ \bibinfo
  {pages} {012001} (\bibinfo {year} {2014})}\BibitemShut {NoStop}%
\end{thebibliography}%

 \end{document}